\documentclass{aa}
\usepackage{graphicx,psfig}
\begin{document}
   \title{An unusual high-ionization nebula in 
     \object{NGC\,55}\thanks{Based on observations collected at the
       European Southern Observatory, La Silla \& Cerro Paranal
       (Chile); Proposal No.: 64.N-0399(A,B), 66.B-0551(A). Based upon
       data obtained from the ESA XMM-{\it Newton} data archive.}  }
   \author{R. T\"ullmann\inst{1} and M.R.
     Rosa\inst{2}\fnmsep\thanks{Affiliated with the Space Telescope
       Division of the European Space Agency, ESTEC, Noordwijk, the
       Netherlands} }

   \offprints{R. T\"ullmann,\\
   \email{tullmann@astro.rub.de}
   }

\authorrunning{T\"ullmann and Rosa}

\institute{Astronomisches Institut, Ruhr-Universit\"at Bochum,
  Universit\"atsstr. 150, D-44780 Bochum, Germany \and Space Telescope
  European Coordinating Facility, c/o European Southern Observatory,
  Karl-Schwarzschild-Str. 2, D-85748 Garching, Germany }
\date{Received 25 August 2003 / Accepted 1 December 2003}

\abstract{We report on the detection of a previously unknown extended
  low-density nebula of very high temperature and excitation,
  TR\,001507.7$-$391206, located above the midplane of NGC\,55. The
  nebular emission line spectrum, in which \ion{He}{ii} is present, is
  consistent with photoionization by about 3 very hot massive O3 or
  WR-type stars. There are no indications for shock ionization. The
  faint blue optical continuum and XMM-{\it Newton} EPIC-pn and
  OM-data also support our assumption that this nebula is not a
  supernova remnant.  Galactic nebulae harboring massive Wolf-Rayet
  (WR) stars may be appropriate examples, in particular the highly
  ionized nebula G2.4+1.4 around WR 102 (aka Sk\,4 or LSS\,4368). The
  relatively large diameter of the inhomogeneously expanding nebula in
  NGC\,55 of about 58\,pc, compared to 11\,pc of G2.4+1.4, would also
  be consistent with the number of ionizing stars.  
   } 
\maketitle
\section{Introduction}
During a study on the Wolf-Rayet (WR) star content of NGC\,55, a
  SBm-type galaxy of the Sculptor-group (e.g., Graham \& Lawrie
  \cite{graham}, Ferguson et al. \cite{fergi}, Otte \& Dettmar
  \cite{otte}, T\"ullmann et al.  \cite{tullmann}) with EFOSC1 at the
ESO 3.6m telescope in 1990, narrow-band images in H$\alpha$ and
[\ion{O}{iii}]$\lambda$5007 revealed a faint, round, featureless
nebular object.  Contrary to the classical giant \ion{H}{ii}-regions
nearby in the body of NGC\,55, it did not show any central peaking of
the nebular emission, nor did any stellar-like central source seem to
be present in medium-band continuum images. One of the longslit
spectra taken for the stellar cores of the giant \ion{H}{ii}-regions
was placed across the nebula to yield a first test spectrum. As
exposed for 1200\,s, the observed spectrum was very weak, but showed
surprisingly strong [\ion{O}{iii}]-lines. Even
[\ion{O}{iii}]$\lambda$4363 was measurable, and yielded a rough
estimate of 20\,000\,K for the electron temperature. This is very high
in comparison to values typical for the giant \ion{H}{ii}-regions in
NGC\,55 of about 11\,500\,K (T\"ullmann et al. \cite{tullmann}).

The object certainly deserved further study, but it was only with more
sensitive equipment that deeper spectra and images could be obtained
in reasonable exposure times. A decade after discovery we now can
benefit from deeper optical spectroscopy, images from the VLT, and
UV/X-ray imaging from the XMM-{\it Newton} satellite.

In the following we present observational results from these
multi-wavelength data and support the analysis with a detailed
modeling of the nebular emission line spectrum in order to identify
the ionization mechanism and to investigate the nature of the central
ionizing source.

\section{The data}
\subsection{Optical spectroscopy and imaging}

The nebula, named TR\,001507.7$-$391206 from now on, was
spectroscopically observed in two different observing runs in June
1990 and October 2000 with ESO's Faint Object and Spectrograph Cameras
EFOSC1 and EFOSC2, attached in both runs to the 3.6m telescope located
at La\,Silla observatory.
\begin{figure*}
\centering
\hspace{-0.75cm}
\includegraphics[width=15.5cm,height=8cm,angle=0]{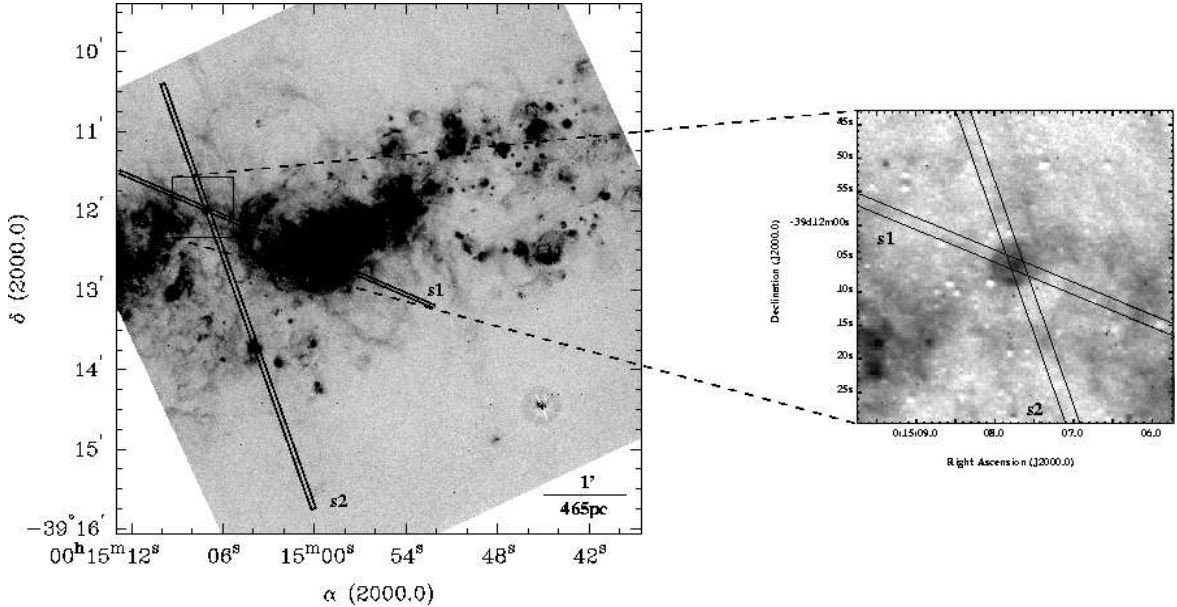}
\caption{This VLT H$\alpha$-image of the central part of NGC\,55 was 
  taken from T\"ullmann et al. (\cite{tullmann}). Slit positions $s1$
  (EFOSC1) and $s2$ (EFOSC2) are scaled to match the chosen slit width
  of 2\arcsec\ and the individual slit lengths of both instruments. The
  magnified image to the right shows the position where both slits cut
  through the nebula.}
\label{fig1}
\end{figure*}
For both spectroscopy runs we selected a rather wide slit of $2\farcs0$
in order to gain maximum flux from the extended target while still
maintaining a reasonable spectral resolution just high enough to make
use of the density sensitive [\ion{S}{ii}] doublet at 6724\AA. Spectra
collected with EFOSC1 in 1990 have a spatial resolution of 0.68\arcsec
pix$^{-1}$, the value of those obtained with EFOSC2 in 2000 amounts to
0.157\arcsec pix$^{-1}$.

During the run in 1990 a number of direct images were obtained with
EFOSC1 through narrow-band filters centered on
[\ion{O}{iii}]$\lambda$5007, [\ion{O}{ii}]$\lambda$3727, H$\alpha$,
H$\beta$, and [\ion{S}{ii}]$\lambda$6727, as well as continuum frames
in medium-band Str\"omgren b, y, and broadband R.  These images
initially led to the discovery of the object but were of mediocre
quality (focus and seeing) to yield any further insight.

In addition, however, we can make use of deeper broadband R and
narrow-band H$\alpha$ VLT-images with much better spatial resolution
obtained during an independent study (T\"ullmann et al.
\cite{tullmann}), covering the same region of the sky. The above
reference also provides information on data reduction techniques for
spectroscopy and photometry which are widely followed here as well. A
journal of observations is presented in Table~\ref{tab1}.

\begin{table}
\caption{Journal of observations}
\begin{tabular}{l l l l l l}
\hline
\hline
\raisebox{.375cm}{}
\hspace{-0.11cm}Telescope & Instrument & Date & Spectral& Duration \\
          &            &           & region             & [ks]     \\
\hline
\raisebox{.375cm}{}
\hspace{-0.11cm}3.6m   & EFOSC1     & $06|20|90$ & [\ion{O}{iii}], V & 0.6, 0.6 \\
          &            &            & 3650--$10^4$\AA   & 1.2      \\
          & EFOSC2     & $10|29|00$ & 3500--$10^4$\AA   & 1.2--3.6 \\
VLT       & FORS1      & $09|12|99$ & H$\alpha$, R      & 0.6, 0.12\\
XMM       & pn         & $11|15|01$ & 0.3--12\,keV      & 28.3     \\
          & OM         & $11|15|01$ & U                 & 0.8      \\
          &            &            & UVW1              & 1.06     \\     
          &            &            & UVW2              & 2.3      \\     
\hline
\end{tabular}
\label{tab1}
\end{table}

In Fig.~\ref{fig1} the VLT H$\alpha$-image is shown together with
detailed positions and dimensions of the two slits $s1$ and $s2$. The
magnified image to the right clearly reveals some substructure and
indicates which parts of the nebula are covered by the longslits.

\subsection{XMM-{\it Newton}-data} 
We also analyzed XMM-{\it Newton} archive data (Obs.-ID.: 0028740101,
Rev.-No.: 0354) of NGC\,55 using primarily EPIC-pn (Str\"uder et al.
(\cite{struder}) and OM-data (Mason et al. \cite{mason}). Data
reduction was done with standard tasks provided by SAS v.5.4.1.

After first inspection of the pn data, a light curve was created to
check for flaring events and to select the good time interval (GTI).
As no background flaring was found to be present, the full exposure
time was usable (Table~\ref{tab1}). The eventlist was filtered
selecting PATTERN $\le12$ and FLAG=0 and cleaned images were produced
in the three energy bands 0.3\,--\,12\,keV (total), 0.3\,--2.0\,keV
(soft) and 2.0\,--\,12.0\,keV (hard).

\begin{figure*}
\includegraphics[width=5.5cm,height=9.cm,angle=-90]{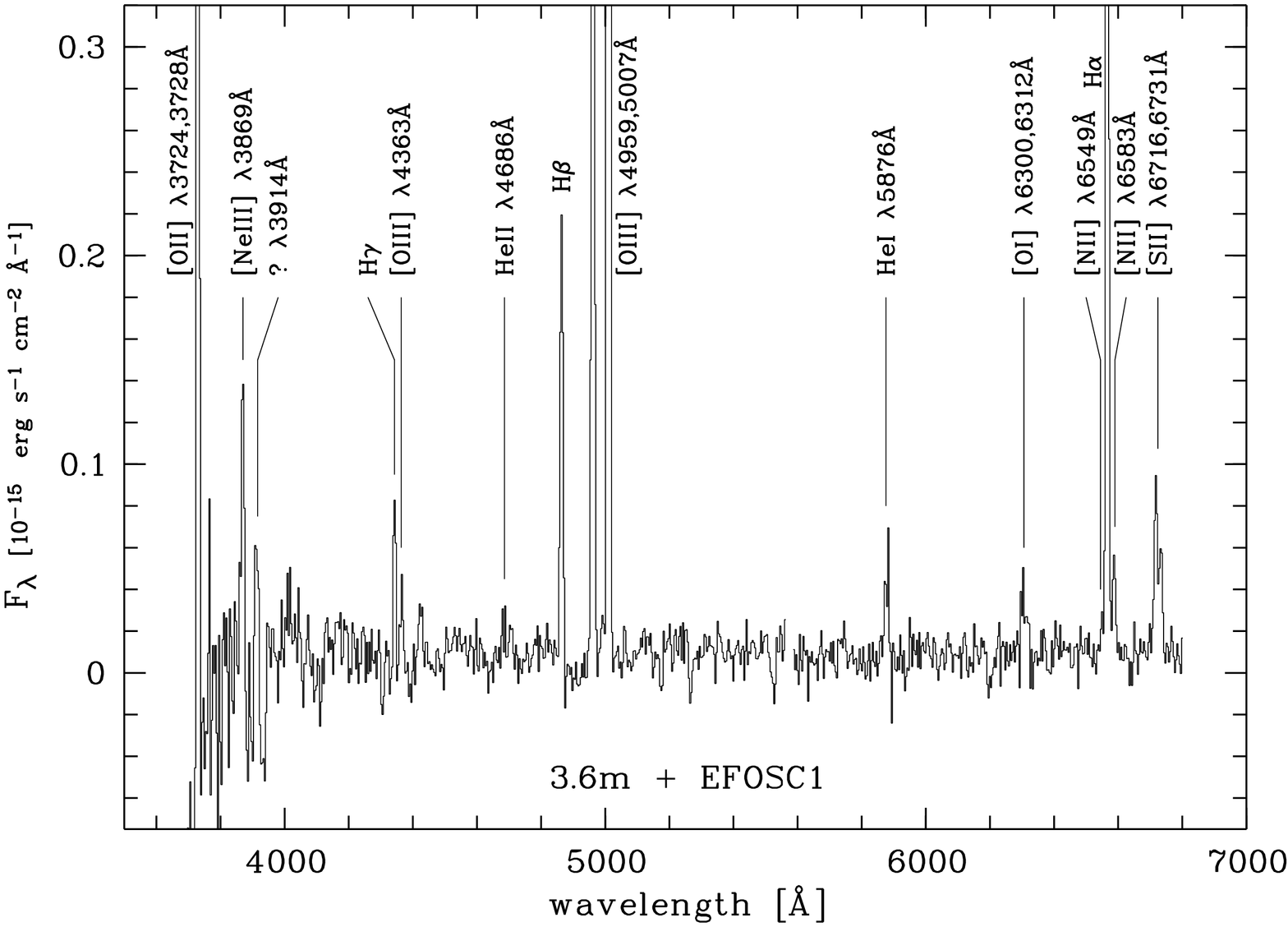}
\hfill
\includegraphics[width=5.5cm,height=9.cm,angle=-90]{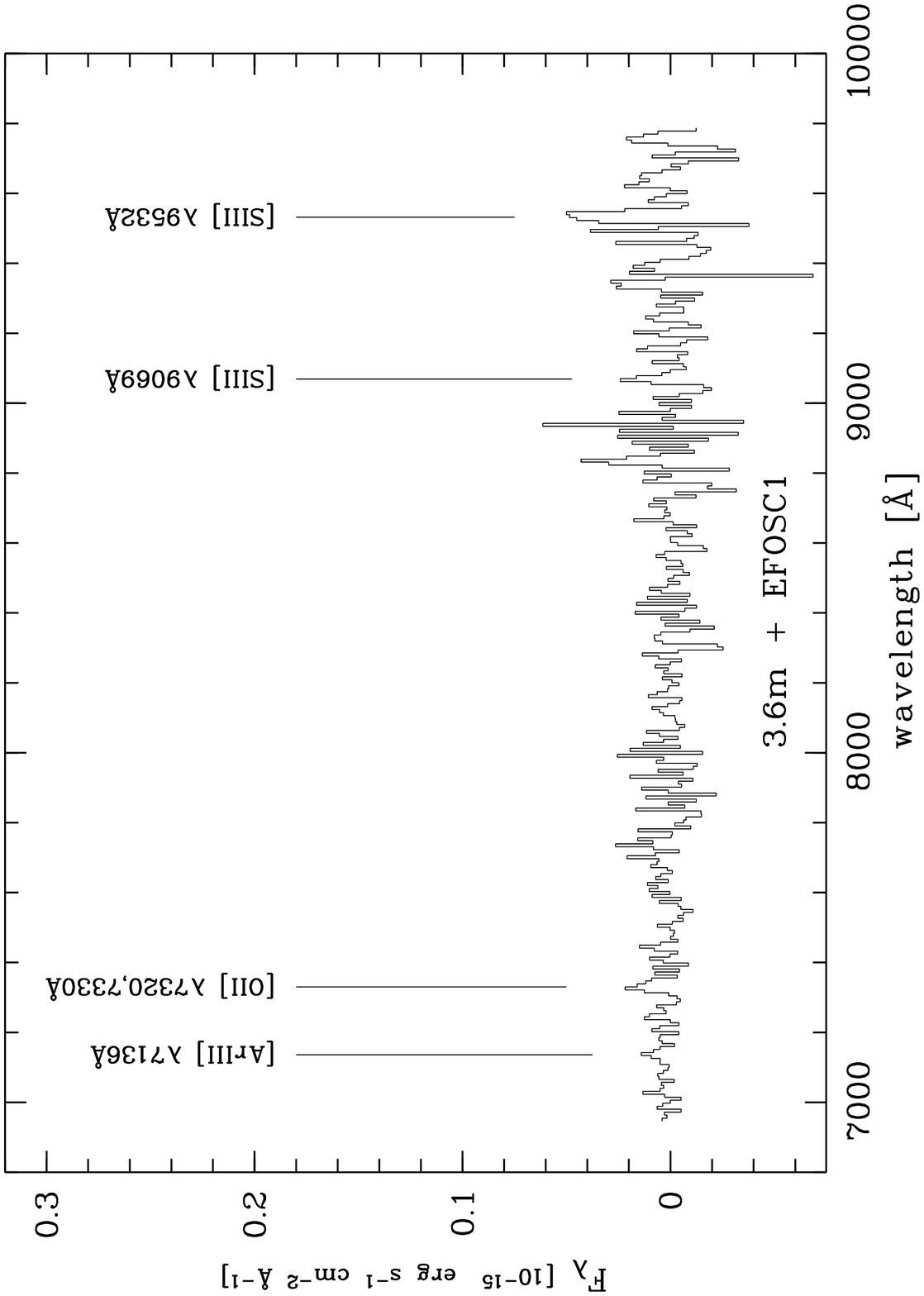}
\caption{Fully-reduced 1D-spectra of TR\,001507.7$-$391206 obtained with 
  EFOSC1 during the first spectroscopy run in June 1990 (slit $s1$).
  The spectra are the sum of the emitted flux from the central part
  and the nebula. Most of the \ion{He}{ii}$\lambda 4686$ line emission
  detected at slit position $s1$ is concentrated in the western and
  the central part of this object.}
\label{fig2}
\end{figure*}
 
\begin{figure*}
\includegraphics[width=5cm,height=9.cm,angle=-90]{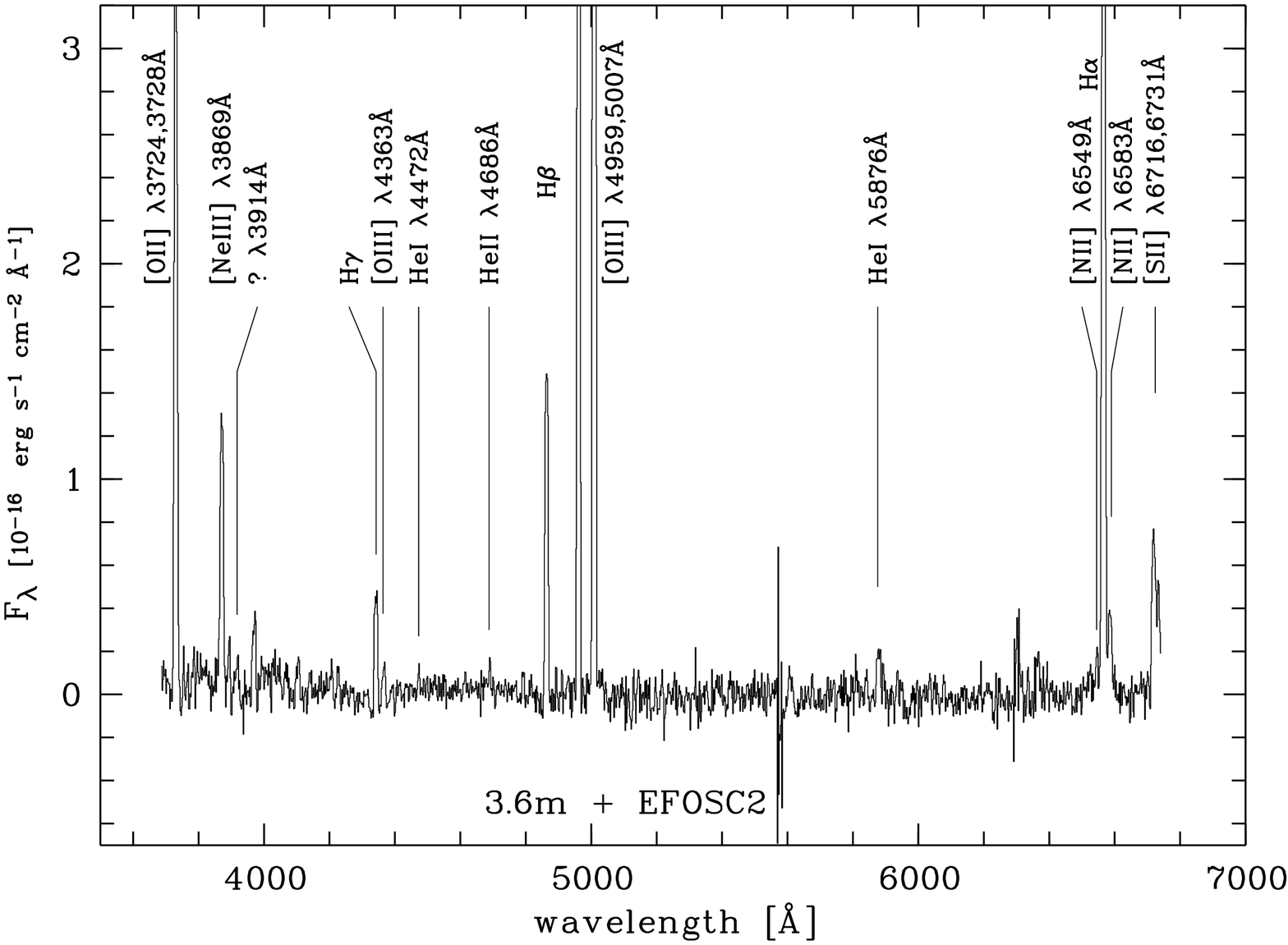}
\hfill
\includegraphics[width=5cm,height=9.cm,angle=-90]{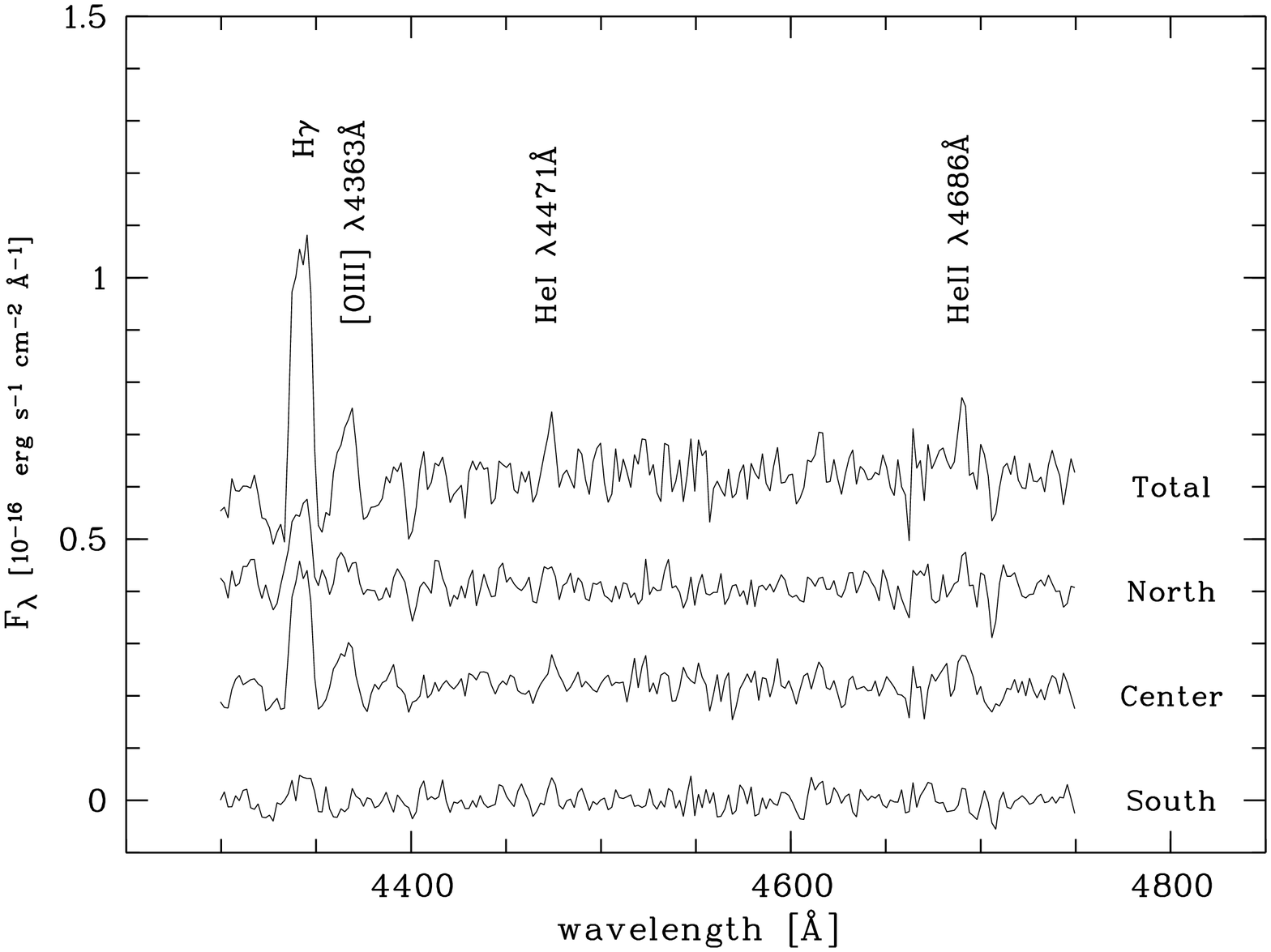}
\caption{This integrated spectrum of slit $s2$ has been obtained with 
  EFOSC2 during the second spectroscopy run in October 2000. Only in
  the center of the nebula very weak stellar continuum emission is
  detectable. Spectral plots, emphasizing \ion{He}{ii}$\lambda
    4686$ and [\ion{O}{iii}]$\lambda 4363$ line emission, are
    displayed with reasonable offsets in the right panel and cut
    through different parts of the nebula (cf. Fig.~\ref{fig1}).}
\label{fig3}
\end{figure*}

OM-data for the UV-filters U, UVW1, and UVW2 were pipeline processed
and calibrated using the SAS-task {\small OMICHAIN}. The {\it PSF
  (FWHM)} of all filters is $\le 2\farcs3$. Integration times for the
corresponding filters are listed in Table~\ref{tab1}. Parasitic light
(e.g., visualized by straylight ellipses, rings, ghosts, etc.) at the
position of the nebula turned out to be negligible.

Due to low count rates, no X-ray spectra could be extracted, even if
all the data collected by the EPIC-cameras (pn and MOS) are combined.

\section{Results and Discussion}
\begin{figure*}
\hspace{.75cm}
\includegraphics[width=17.1cm,height=15.4cm,angle=0]{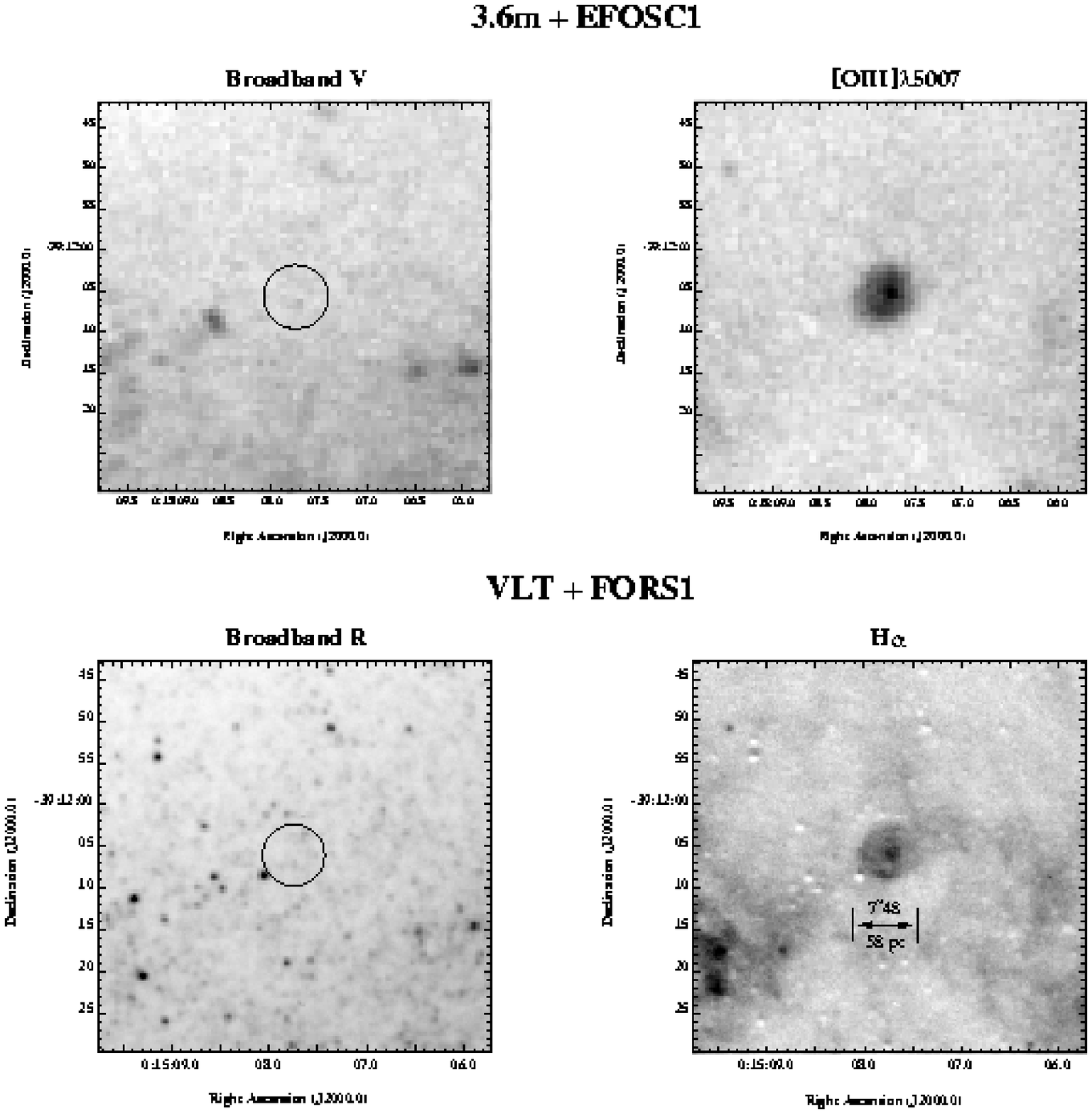}
\caption{Broad- and narrow-band filter images of TR\,001507.7$-$391206 
  obtained with different ESO instruments. Due to better spatial
  resolution of the FORS1 instrument, the VLT H$\alpha$-image reveals
  a central dense core and a fragmented outer gas shell. There seems
  to be very faint continuum emission in the R-band filter image,
  which appears to be slightly stronger in the V-band. For V- and
  R-band photometry the location of the object is marked by a circle.
  The diameter of the nebula has been determined to be 7\farcs5 which
  corresponds to 58\,pc.}
\label{fig4}
\end{figure*}
\begin{figure*}[!t]
\hspace{1.5cm}
\includegraphics[width=16cm,height=20.8cm,angle=0]{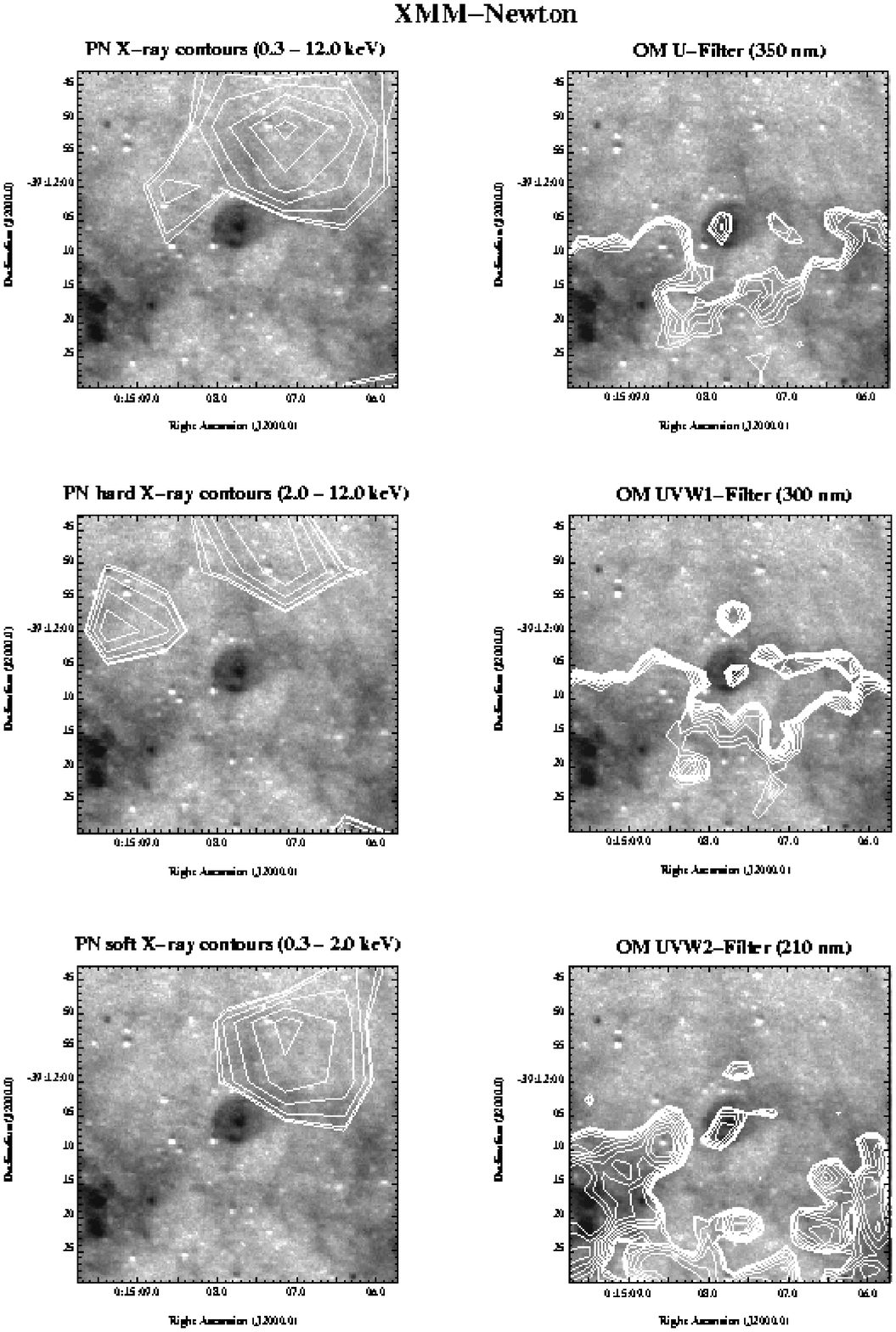}
\caption{XMM-{\it Newton} view of the same region. X-ray contours (EPIC-pn 
  data, left panel) for different energies are overplotted on the VLT
  H$\alpha$-image. In addition, UV broadband contours from the optical
  monitor (OM, right panel) unambiguously reveal continuum emission
  indicating most likely the presence of a hot stellar source in the
  center of the nebula.}
\label{fig5}
\end{figure*}
If projection effects along the line of sight are irrelevant,
TR\,001507.7$-$391206 is located on top of an extended V-shaped column
of gas and dust (seen upside down), that is protruding off the disk
plane of NGC\,55. This is shown very nicely in Fig.~\ref{fig1}, where
the gaseous structure is aligned with slit position $s2$.

A spectrum of the nebula covering the wavelength range from
3500\AA\,--\,10\,000\AA\ is presented for slit $s1$ in Fig.~\ref{fig2}.
Since the airglow was much stronger during the EFOSC2 run, the spectrum
for slit location $s2$ is displayed only in the range of
3500\AA\,--\,7000\AA\ (Fig.~\ref{fig3}, left panel).

In Fig.~\ref{fig3} (right panel) we also show spectra from slit position
$s2$ sampling different sections of the slit.  These plots reveal that
the total integrated emission of \ion{He}{ii}\,$\lambda$4686 for slit
$s2$ is weak, but significant.  Apparently, this emission is
concentrated in the northern and central parts of the nebula, whereas
there is no clear detection for the southern part. The co-added data
for slit $s1$ are of insufficient S/N to investigate spatial
variations of the faint \ion{He}{ii}\,$\lambda$4686 line emission.

Although this emission line appears to be relatively broad at slit
position $s2$, it is completely unresolved, as the FWHM is only half
of that of H$\gamma$ or [\ion{O}{iii}]$\lambda$4363. Therefore, line
broadening effects (e.g., due to stellar winds) are not supported by
the data presented here. Since the \ion{He}{ii}\,$\lambda$4686
emission is spatially extended, and the stellar continuum is extremely
weak, it is safe to assume that all of the emission is nebular in
origin (see also flux estimates below).

The spectrum of the red wavelength region obtained at slit position
$s1$ does not show stellar continuum emission at all.  However, there
is very faint continuum detectable within the central 15\,pc at slit
position $s2$ throughout the whole covered wavelength region.  In
addition, continuum is most likely present in the V-band image and, to
a significantly lower extent, also on the R-band frame (cf.
Fig.~\ref{fig4}).  Among the data presented here, this emission
appears to be strongest on the UV broadband images obtained with the
OM-telescope of XMM-{\it Newton} (Fig.~\ref{fig5}).  The presence of a
weak continuum, slightly stronger in the blue, is confirmed by the
spectra shown in Fig.~\ref{fig2}.
 
Although the spectrum of the potential central ionizing source is too
weak to be extracted, we can constrain its nature indirectly by
analyzing the ionization mechanism and measuring ionic and elemental
abundances of the surrounding nebula.

Observed and dereddened emission line fluxes are listed in
Table~\ref{tab2} together with predictions of the nebular abundance
tool (NAT, see text below).  Because of the high (negative) galactic
latitude of NGC\,55 almost all of the reddening is due to dust located
in that galaxy. In the absence of a dedicated reddening law valid for
NGC\,55 we used the standard galactic law for dereddening. Even if we
had used the optical part of the SMC-law the reported line ratios
would not have changed significantly for the present purpose.

\begin{table}
\caption{Dereddened emission line fluxes for both slit positions listed 
together with corresponding NAT predictions. Since the spectral resolution 
is too low to resolve $[\ion{O}{i}]$\,$\lambda$6300 and 
$[\ion{S}{iii}]$\,$\lambda$6311, the sum of both lines is given instead. 
The same holds for the oxygen-doublets at 3727\AA\ and 7325\AA.}
\begin{tabular}{l c c c c}
\hline
\hline
\raisebox{.375cm}{}
\hspace{-.11cm}Line & slit\,1 & NAT & slit\,2 & NAT \\
H$\beta = 100$ & & & & \\
\hline
\raisebox{.375cm}{}
\hspace{-.1cm}$\Sigma\ [\ion{O}{ii}]$\,$\lambda$3727 & 315 & 315 & 499 & 500  \\
$[\ion{Ne}{iii}]$\,$\lambda$3869       & 83.9 & 83.9 & 156  & 156  \\
H$\gamma$                              & 34.3 & 47.5 & 46.9 & 47.5 \\
$[\ion{O}{iii}]$\,$\lambda$4363        & 23.3 & 23.2 & 19.6 & 19.3 \\
$\ion{He}{i}$\,$\lambda$4472           & --   & 10.3 & 5.62 & 6.49 \\
$\ion{He}{ii}$\,$\lambda$4686          & 4.88 & --   & 2.60 & --   \\
H$\beta$                               & 100  & 100  & 100  & 100  \\
$[\ion{O}{iii}]$\,$\lambda$4959        & 276  & 279  & 227  & 233  \\
$[\ion{O}{iii}]$\,$\lambda$5007        & 814  & 805  & 686  & 670  \\
$[\ion{Ar}{iii}]$\,$\lambda$5193       & --   & 0.24 & 2.42 & --   \\
$[\ion{Cl}{iii}]$\,$\lambda$5518       & 9.54 & 9.54 & 0.88 & 0.88 \\
\ion{He}{i}\,$\lambda$5876             & 27.4 & 27.4 & 19.5 & 17.2 \\
$
\left .
\begin{tabular}{l}
\hspace{-.255cm}$[\ion{O}{i}]$\,$\lambda$6300\hspace{-.2cm} \\         
\hspace{-.255cm}$[\ion{S}{iii}]$\,$\lambda$6311\hspace{-.2cm} 
\end{tabular} 
\right\} 
$                                      & 36.1 & 4.42 & --   & --  \\
$[\ion{N}{ii}]$\,$\lambda$6548         & 5.05 & 5.83 & 3.81 & 6.43\\
H$\alpha$                              & 274  & 274  & 274  & 274 \\
$[\ion{N}{ii}]$\,$\lambda$6583         & 19.5 & 17.2 & 26.7 & 18.9\\
$[\ion{S}{ii}]$\,$\lambda$6716         & 40.5 & 36.0 & 54.2 & 48.5\\
$[\ion{S}{ii}]$\,$\lambda$6731         & 22.1 & 25.3 & 30.0 & 34.1\\
$[\ion{Ar}{iii}]$\,$\lambda$7136       & 11.3 & 11.3 & --   & --  \\
$\Sigma\ [\ion{O}{ii}]$\,$\lambda$7325 & 23.8 & 5.09 & --   & 4.30\\
$[\ion{S}{iii}]$\,$\lambda$9072        & 28.2 & 27.7 & --   & --  \\
$[\ion{S}{iii}]$\,$\lambda$9535        & 67.7 & 68.8 & --   & --  \\
\hline 
\raisebox{.375cm}{}
\hspace{-0.11cm}$c_{\rm ext}$                     & 0.45         & --       & 0.45     & --      \\
$T_{\rm e}$([\ion{O}{iii}]) [K]   & 18\,300      & 18\,300  & 18\,300  & 18\,300 \\
$n_{\rm e}$ [cm$^{-3}$]             & $\le 10$     & $\le 10$ & $\le 10$ & $\le 10$\\
$v_{\rm hel}$ [km s$^{-1}$]       & $161 \pm 23$ & --       & $157 \pm 27$ & --  \\
\hline
\end{tabular}
\label{tab2}
\end{table}

\subsection{Gas phase abundances}
The direct detection of $[\ion{O}{iii}]$\,$\lambda$4363 enables us to
obtain a reliable estimate of the gas temperature and, together with
constraints on the density, ionic abundances of the gas. Abundances
are determined with NAT, a code based on the 5-level-atom program of
De Robertis et al. (\cite{derobert}) (see T\"ullmann et al.
\cite{tullmann} for details). If only a single transition of an
element could be observed (e.g., [\ion{Ne}{iii}] or [\ion{Ar}{iii}]),
NAT adopts this value to calculate fractional ionizations and element
abundances. However, since only photon energies $<54$\,eV are
considered, no ionic abundances are calculated for \ion{He}{ii} by
this tool.

A final conversion from ionic to element abundances is achieved by
applying the ICF scheme of Mathis \& Rosa (\cite{maro}) which allows
us to use parameterized results from photoionization models with a wide
variety of stellar atmosphere models. Element abundances determined
this way are given in Table~\ref{tab3}. For convenience, we consider
the metallicity $Z$ to be equal to $\log({\rm O/H})$, since oxygen is
the most abundant and efficient coolant.

In relation to the solar values (Christensen-Dalsgaard
\cite{christensen}, Grevesse \& Sauval \cite{grevesse}),
[He$^+$/H$^+$] appears to be overabundant by about a factor of 2.5 at
slit position $s1$ and by about a factor of 1.5 at position $s2$.  The
total atomic He/H ratio will be slightly higher if the ionization of
\ion{He}{ii} is accounted for. [Ne/O] at position $s2$ appears to be
overabundant by about the same value of 1.5.  However, Ne-abundances
are based on only a single Ne$^{++}$ transition at 3869\AA\ and are
therefore less certain.  At both slit positions the [S/O]-ratio also
turns out to be enhanced by an average factor of 1.3.
\begin{table}
\caption{Derived abundances (with representative errors for slit $s1$) 
of the candidate WR nebula compared to averaged abundances of both EHRs 
and the central \ion{H}{ii}-region in NGC\,55 (T\"ullmann et al. 
\cite{tullmann}).}
\begin{tabular}{l r r r r}
\hline
\hline
\raisebox{.375cm}{}
\hspace{-0.09cm}Parameter & slit\,$s1$ & slit\,$s2$ & EHRs & HR\\
\hline
\raisebox{.375cm}{}
\hspace{-0.11cm}$12+\log({\rm He/H})$&$11.36\pm 0.05$&$11.16$&10.90&10.94   \\
$12+\log({\rm O/H})$       & $7.84\pm 0.08$  & $7.83$  & 7.72     & 8.05    \\
$\log({\rm N/O})$          & $-1.35\pm 0.05$ & $-1.44$ & $-$1.45  & $-$1.26 \\
$\log({\rm Ne/O})$         & $-0.89\pm 0.15$ & $-0.49$ & ---      & $-$0.85 \\
$\log({\rm S/O})$          & $-1.40\pm 0.10$ & $-1.38$ & $-$1.80  & $-$1.41 \\
$\log({\rm O^{+}/O})$      & $-$0.671        & $-$0.469& $-$0.130 & $-$0.731\\
$\log({\rm S^{+}/S^{++}})$ & $-$0.503        & ---     & ---      & $-$0.722\\
$\overline{Z/Z_{\odot}}$   & 0.13            & 0.13    & 0.11     & 0.23    \\
\hline
\end{tabular}
\label{tab3}
\end{table}

The ionization of the nebula is obviously rather high. At slit
position $s1$ ($s2$) only $21\%$ ($34\%$) of oxygen is expected to be
present as O$^{+}$ and $79\%$ ($66\%$) as O$^{++}$. Therefore, it is
remarkable that the oxygen abundance of TR\,001507.7$-$391206 is
within 0.1\,dex identical to that found for the extraplanar
\ion{H}{ii}-regions (EHRs) in the disk-halo interface of NGC\,55
(T\"ullmann et al. \cite{tullmann}). Similar to the EHRs, the nebula
is also located above the central star forming complex, which would
imply that the EHRs and the nebula have been formed from the same gas
that has been pushed into the halo by the combined effects of stellar
winds and supernova explosions during an earlier epoch of star
formation in the galactic disk of NGC\,55.

\subsection{The nature of the nebula}
The detection of nebular \ion{He}{ii}$\lambda 4686$ emission in both
spectra puts some constraints on the ionizing source of the nebula.
Since 54.4\,eV are required to ionize helium completely, only the
hottest O stars or their evolutionary products (WR stars) provide
sufficient luminosities at these photon energies.  However,
non-thermal ionizing mechanisms, such as in supernova remnants (SNRs),
are not a priori excluded at this point.

Another constraint which also implies a hot atmosphere, if
photoionization is chosen, comes from the faintness of any probable
stellar continuum in the ``red'' wavelength region, which seems to
become stronger towards the ``blue'', and apparently is strongest on
UV broadband data collected with the optical monitor (OM) of the
XMM-{\it Newton} satellite.

Further support of an extreme ionizing source is given by the high gas
temperature of about 18\,300\,K at a metal abundance which is
otherwise comparable to the range of abundances observed for other
nebulae in NGC\,55. These \ion{H}{ii}-regions have typical gas
temperatures of only 11\,500\,K and oxygen abundances of about
$10\%\,Z_{\odot}$ (T\"ullmann et al. \cite{tullmann}).

XMM-{\it Newton} EPIC-pn data obtained at different energy bands and
overlaid onto the VLT H$\alpha$-image (Fig.~\ref{fig5}) reveal a
significant detection of hard and soft X-rays around the northern part
of the nebula. This emission is too well aligned with the outer
optical border of the nebula to be just a normal line-of-sight effect.
The questions if the lack of X-ray photons to the south of the nebula
is caused by absorption of the ambient ISM (see Fig.~\ref{fig1}) and
if the extended X-ray emission in the north is related to the star
formation activity in the disk below, remains yet to be answered.

In principle shock ionization could also be a possible ionization
mechanism of the gas in the nebula under discussion.  However, the
observed emission line characteristics clearly indicate that the nebula
is likely photoionized rather than shock ionized: in particular the
weakness of [\ion{S}{ii}], [\ion{S}{iii}], and [\ion{N}{ii}] emission
lines; the absence of \ion{Mg}{ii} emission and the weakness of
[\ion{O}{i}]$\lambda$6300. Diagnostic diagrams such as those from
Baldwin et al. (\cite{bald}) or Veilleux \& Osterbrock (\cite{veil}), and
a comparison with shock models (Shull \& McKee \cite{shull}, Raymond
\cite{ray}, Dopita \& Sutherland \cite{dosu}) also indicate
photoionization conditions.

In order to check the hypothesis that the dominant ionizing
sources are massive stars, observed emission line strengths are
compared to predictions made by NAT in an iterative process. The best
match between model and observations was achieved using a model
atmosphere for a WR star. In this case a pure-helium extended atmosphere 
of $T=90\,000$\,K (Wessolowski et al. \cite{wesso}) was
chosen. Results are shown in Table~\ref{tab2} in the columns labeled
``NAT''. They are in very good agreement with the observations and
support the idea that TR\,001507.7$-$391206 is photoionized by evolved
massive stars.

Since NAT cannot predict \ion{He}{ii}/H line ratios, we now estimate the
theoretical \ion{He}{ii}/H$\beta$ line ratio and comparing it to the
observed one (cf. Table~\ref{tab2}).  Following Osterbrock
(\cite{oster}), this ratio can be written as:
\begin{equation}
\frac{\ion{He}{ii}\lambda 4686}{{\rm H}\beta}\approx \frac{\int N_{{\rm He}^{++}}N_{{\rm e}}\ \alpha_{\lambda 4686}^{{\rm eff}}({\rm He}^{+},T)\ dV_{{\rm He}^{++}}}
{\int N_{{\rm p}}N_{{\rm e}}\ \alpha_{{\rm H}\beta}^{{\rm eff}}(H^0,T)\ dV_{{\rm H}^+}},
\end{equation}
where $N_{{\rm He}^{++}}, N_{{\rm p}}$, and $N_{{\rm e}}$ represent
ion, proton, and electron densities, respectively. $\alpha^{{\rm
eff}}$ denotes the effective recombination coefficient of the
corresponding transition whereas $dV$ addresses the respective volume
of the H$^{+}$ and He$^{++}$-zone covered by the slit.

We assume that H is completely ionized and that $r_{{\rm He}^{++}}$,
the radius of the He$^{++}$-zone, is about $0.2\,r_{{\rm H}^+}$. This
is reasonable for nebulae ionized by stars with effective temperatures
of the order of $10^5$\,K, e.g., planetary nebulae. Then $V_{{\rm
    He}^{++}}/V_{{\rm H}^{+}}$ amounts to 0.04.  Using tabulated
recombination coefficients for \ion{He}{ii} and H$\beta$ (Osterbrock
\cite{oster}, Tables~4.2 and 4.3) and the approximation that for high
ionization nebulae $N_{{\rm He}^{++}}/N_{{\rm H}^{+}}\approx N_{\rm
  He}/N_{\rm H}\approx 0.1$, we finally derive: $\ion{He}{ii}\lambda
4686/{\rm H}\beta\approx 0.044$. As this is in good agreement with
observational data, it appears to be justified to consider very hot
massive stars, such as O3 or WR stars, to be the ionizing sources of
TR\,001507.7$-$391206.

We now derive an estimate of the number of such ionizing stars by
converting the integrated H$\alpha$-flux into a luminosity, assuming a
distance to NGC\,55 of $D=1.6$\,Mpc, and comparing this value to what
is expected from theory. For slit $s1$ the integrated H$\alpha$-flux
amounts to $F_{\rm H\alpha,s1}=1.54\times 10^{-13}$\,erg
s$^{-1}$\,cm$^{-2}$. At slit position $s2$ we measure a value for
$F_{\rm H\alpha,s2}$ of $2.26\times 10^{-13}$\,erg
s$^{-1}$\,cm$^{-2}$. This translates into the following luminosities:
$L_{\rm H\alpha,s1}=4.7\times 10^{37}$\,erg s$^{-1}$ and $L_{\rm
  H\alpha,s2}=6.9\times 10^{37}$\,erg s$^{-1}$.  Since the
NAT-modeling of the observed emission line data yielded a typical
effective temperature of 90\,000\,K at a metallicity of 10$\%$ solar,
such stars are expected to have a photon luminosity of the \ion{H}{i}
ionizing continuum of about $1.58\times 10^{49}$\,s$^{-1}$ (Smith et
al. \cite{smith}).  Recombination line luminosities are calculated
using case B recombination coefficients for H as given in Table~2.1 of
Osterbrock (\cite{oster}). From this data we derive expected stellar
photon luminosities of $L_{\rm H\alpha}=5.6\times
10^{37}$\,erg s$^{-1}$. If we scale this by 0.39, the fraction of the
nebular volume which each slit covers, we finally derive a corrected
luminosity of $L_{\rm H\alpha}=2.2\times 10^{37}$\,erg s$^{-1}$. This
would be consistent with observed H$\alpha$-luminosities if the
central ionizing source consists of two to three WR stars.

It would be very interesting to obtain a much deeper spectrum of the
central continuum sources in order to investigate the nature of these
stars in detail. The presence of WR-typical broad emission lines as
well as absorption features at wavelengths where helium and hydrogen
lines do not coincide (e.g., at 4200\AA\ or 4542\AA), would be
indicators for the mix of early O and WR stars.

\begin{figure}
\includegraphics[width=5.5cm,height=8.8cm,angle=-90]{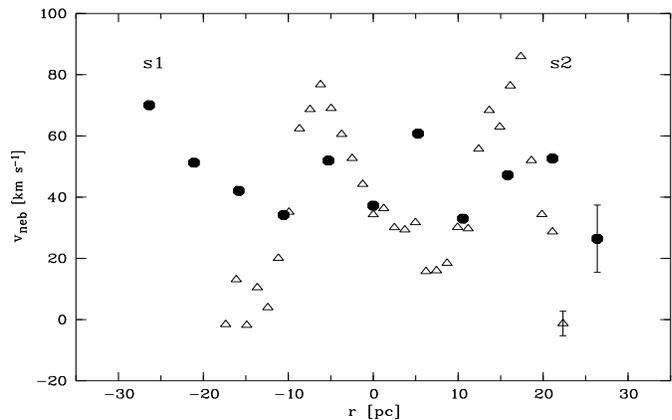}
\caption{Nebular H$\alpha$-velocities as a function of radius $r$ plotted 
for both slit positions. A correction for the systemic velocity has
been applied, assuming $v_{{\rm sys}}= 118\,{\rm km}\ {\rm s}^{-1}$
(Puche et al. \cite{puche}). Typical uncertainties in velocity are in
the range of $4-12$\,km s$^{-1}$ (see errorbars).  Negative radial
coordinates for $s1$ ($s2$) represent the western (southern) part of
the nebula.}
\label{fig6}
\end{figure}

\subsection{Nebular kinematics}
Measurements of radial velocities determined from slit $s2$,
10\arcsec\ south of TR\,001507.7$-$391206 (dust cloud), reveal values
ranging from $-$1\,km s$^{-1}$ to $+45$\,km s$^{-1}$. As this range is
identical to velocities found for the nebula (Fig.~\ref{fig6}), it
implies that this object is indeed located on top of a huge gas spike.

Data for $s2$ show that the ionized gas at the outskirts of the nebula
meets the systemic velocity (Fig.~\ref{fig6}) and is therefore mostly
at rest. Hence these regions might be coincident with the outer
(stationary) border of the nebula. At $s1$ the kinematics looks
completely different and the systemic velocity is never approached. 
This tells at least that TR\,001507.7$-$391206 is not expanding
symmetrically.

\subsection{Comparison with other objects}
Although direct observational evidence of the exact stellar types is
lacking in case of TR\,001507.7$-$391206, the overabundance of He, S, and
possibly Ne as well as the observed range of velocities seem to imply that we
are faced with a mass-loss bubble very much like those around galactic
WR stars.  Since the stellar ionizing sources for
TR\,001507.7$-$391206 are apparently very hot, a suitable candidate
for comparison would be the galactic nebula G2.4+1.4.  This
extraordinary mass-loss bubble around WR\,102, a WO-type WR star, has
been described in detail by e.g. Treffers \& Chu (\cite{treffer}),
Dopita et al. (\cite{dopita1}), Dopita \& Lozinskaia (\cite{dopita2}),
or Esteban et al. (\cite{esteban}). WR\,102 is presumably the hottest
and most evolved massive WR star known in the Galaxy.  A comparison
between H$\alpha$-images of G2.4+1.4 and TR\,001507.7$-$391206 reveals
remarkable morphological similarities if this object would be
projected to the distance of NGC\,55.

If the equivalences between the WR nebulae in NGC\,55 and the
Galaxy hold true then TR\,001507.7$-$391206 harbors one of the most
massive and short-lived stars of NGC\,55. 
Clearly, deep exposures using the VLT would be required to obtain spectra of 
the central sources with sufficient signal to noise.

\section{Summary}
We have collected multi-wavelength data in order to constrain the
nature of a high ionization, high temperature nebula located in the
diffuse ISM near the main body of NGC\,55. Its emission line
characteristics, gas temperature, and elevated [He/H], [S/O], and possibly
[Ne/O]-abundance are consistent with slightly enriched, plowed matter
present either in SNRs or WR nebulae and ionized by a hot
continuum. Optical and UV broadband data reveal the existence of
continuum emission originating from this object which directly
excludes SNRs. A SNR nature is also made improbable, because the
characteristic emission of \ion{Mg}{ii} and [\ion{O}{i}]$\lambda$6300
is absent. Moreover, observed emission line intensities compare well
with photoionization models using a hot WR atmosphere of
$T=90\,000$\,K as input.

The most plausible interpretation of the data is a WR nebula
photoionized by one of the hottest and most massive WR stars (WO)
similar to the nebula G2.4+1.4 in the Galaxy which is ionized by WR
102. Such stars are rather inconspicuous in optical imaging, but very
deep longslit spectra with the VLT across the nebula might well reveal
the characteristic broad emission lines of the central WR object.

\begin{acknowledgements}
RT received financial support by the SFB 591 and DLR through grant
50OR0102. The referee C. Leitherer pointed out some aspects which made
the paper more conclusive.
\end{acknowledgements}

\end{document}